\documentclass[11pt,prd,aps,floatfix,nofootinbib,subeqn]{revtex4-1}

\usepackage{graphicx}
\usepackage{subfigure}
\usepackage{amsmath, amsthm, amssymb,textcomp,esint}
\usepackage{subfigure}
\usepackage{hyperref}
\usepackage[all]{hypcap}

\numberwithin{equation}{section}

\begin{document}

\vspace*{-10mm}
\begin{flushright}
\normalsize
UTHEP-118\\
(Dec. 1983)
\end{flushright}

\title{
Renormalization Group Analysis of Lattice Theories \\and \\Improved Lattice Action. II\\
--- four-dimensional non-abelian SU(N) gauge model ---
}

\author{Y. Iwasaki}

\affiliation{
Institute of Physics, University of Tsukuba\\ Ibaraki 305, Japan
}


\begin{abstract}
Abstract: A new block spin renormalization group transformation for SU(N) gauge models is proposed near the non-trivial fixed point in perturbation theory and 
thereby the expectation values of various Wilson loops on the renormalized trajectory near the fixed point are explicitly obtained. 
An improved action is obtained as in a preceding paper and a criterion for the scaling behavior of physical quantities is also given.
\end{abstract}


\maketitle

\clearpage

\section{Introduction}
\label{section1}
In a preceding paper \cite{Iwasaki-prec} (to be referred as [I] ) we have emphasized the significance to improve a lattice action for the continuum limit and 
have shown our strategy to do it in renormalization group approach for asymptotically free lattice theories such as 2d nonlinear O(N) sigma models and 
4d SU(N) gauge models, most attention has been paid to the former. In particular, the correlation function on the renormalized trajectory has been explicitly 
obtained near the non-trivial fixed point by block spin renormalization group in perturbation theory. An improved action has been obtained from the criterion 
that it is most close to the renormalized trajectory by defining a distance from an action to the renormalized trajectory. A criterion has been also given 
for the scaling behavior of physical quantities obtained by Monte Carlo (MC) simulations with a given lattice action.

In this paper we will make a similar analysis for 4d SU(N) gauge models. We will not repeat in principle the general discussions given already in [I] to avoid 
the repetition. Therefore we assume that the reader is familiar with [I].

The organization of this paper is parallel to that of [I], except for that the sections which correspond to sections 8 and 9 of [I] do not exist in this paper: 
The model is defined in section \ref{section2}, block spin transformation is introduced in section \ref{section3}, the expectation values of various Wilson loops 
on the renormalized trajectory are obtained in section \ref{section4} and the block spin transformation is performed for various lattice actions in section 
\ref{section5}. An improved action is obtained in section \ref{section6}, implications of our results for MC calculations are discussed in section \ref{section7} 
and the connection between the existence of instantons on the lattice and the renormalized trajectory is given in section \ref{section8}. 
Section \ref{section9} is devoted to discussion. In the appendix \ref{appendixA} the explicit form of the propagator 
is given and in the appendix \ref{appendixB} the derivation of eq.(\ref{eq.3-9}) is given.

\section{SU(N) lattice gauge model \cite{Wilson1974} and weak coupling expansion}
\label{section2}
\begin{figure}[t]
\centering
\hspace{-15mm}
\subfigure[]{\includegraphics[scale=0.251]{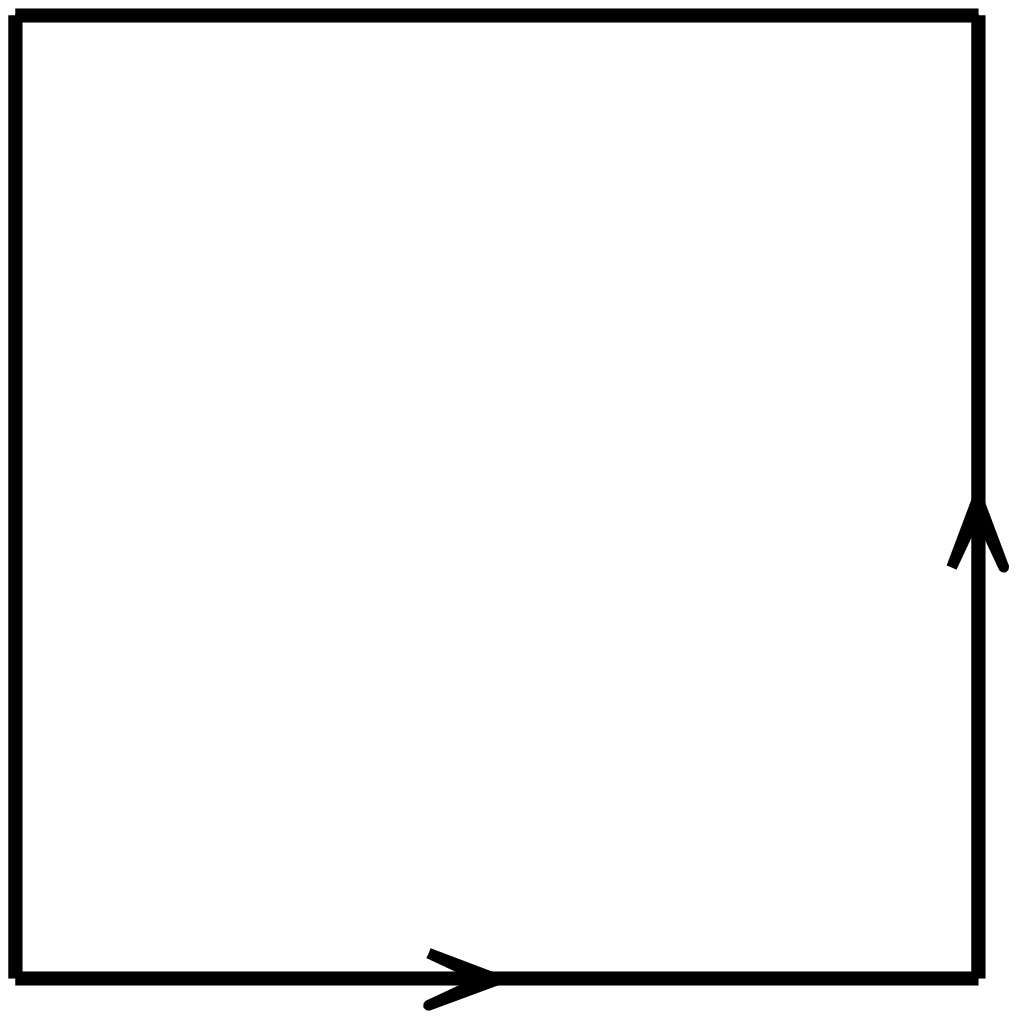}}
\hspace{-15mm}
\subfigure[]{\includegraphics[scale=0.251]{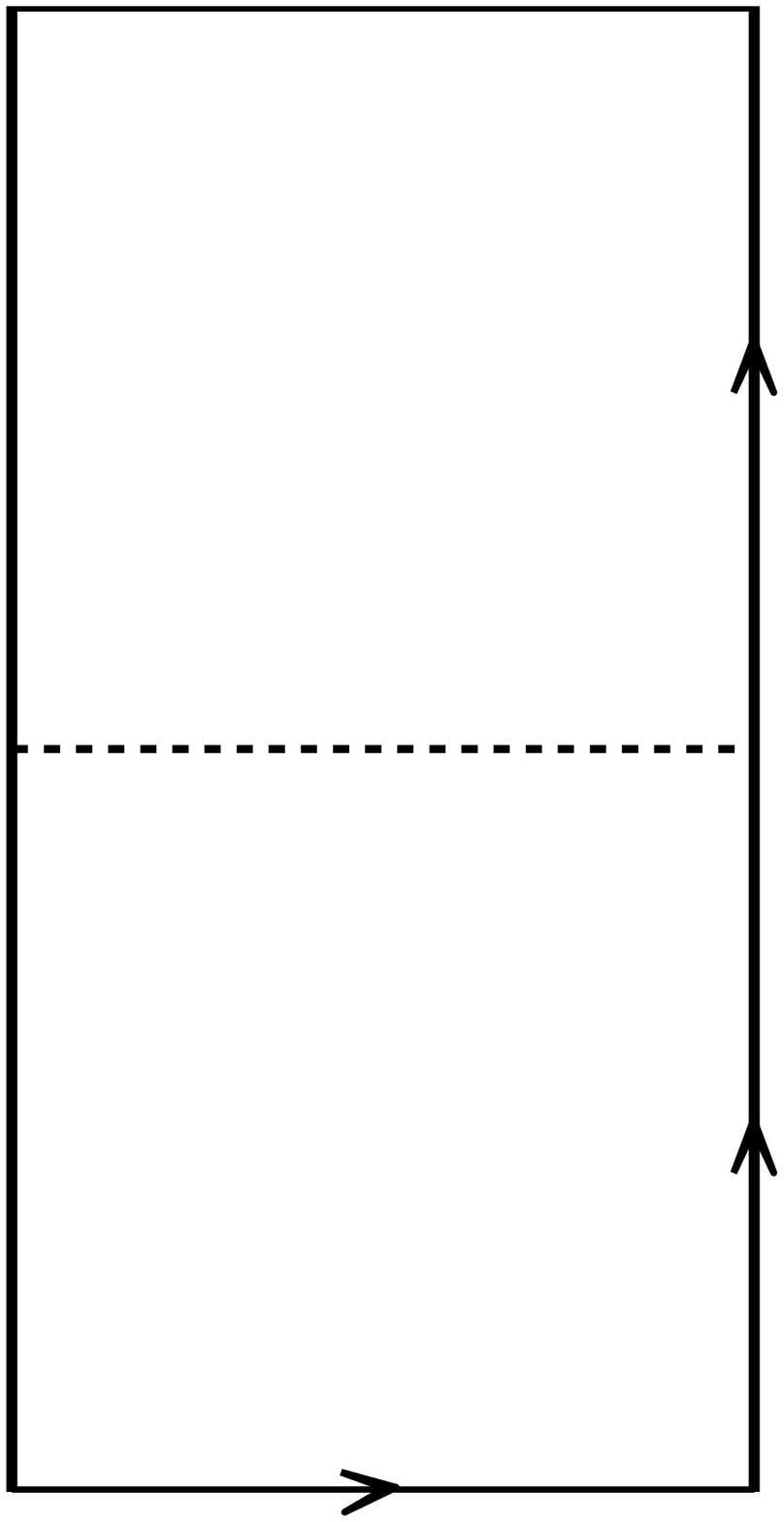}}
\hspace{-15mm}
\\
\vspace{-2mm}
\hspace{-25mm}
\subfigure[]{\includegraphics[scale=0.251]{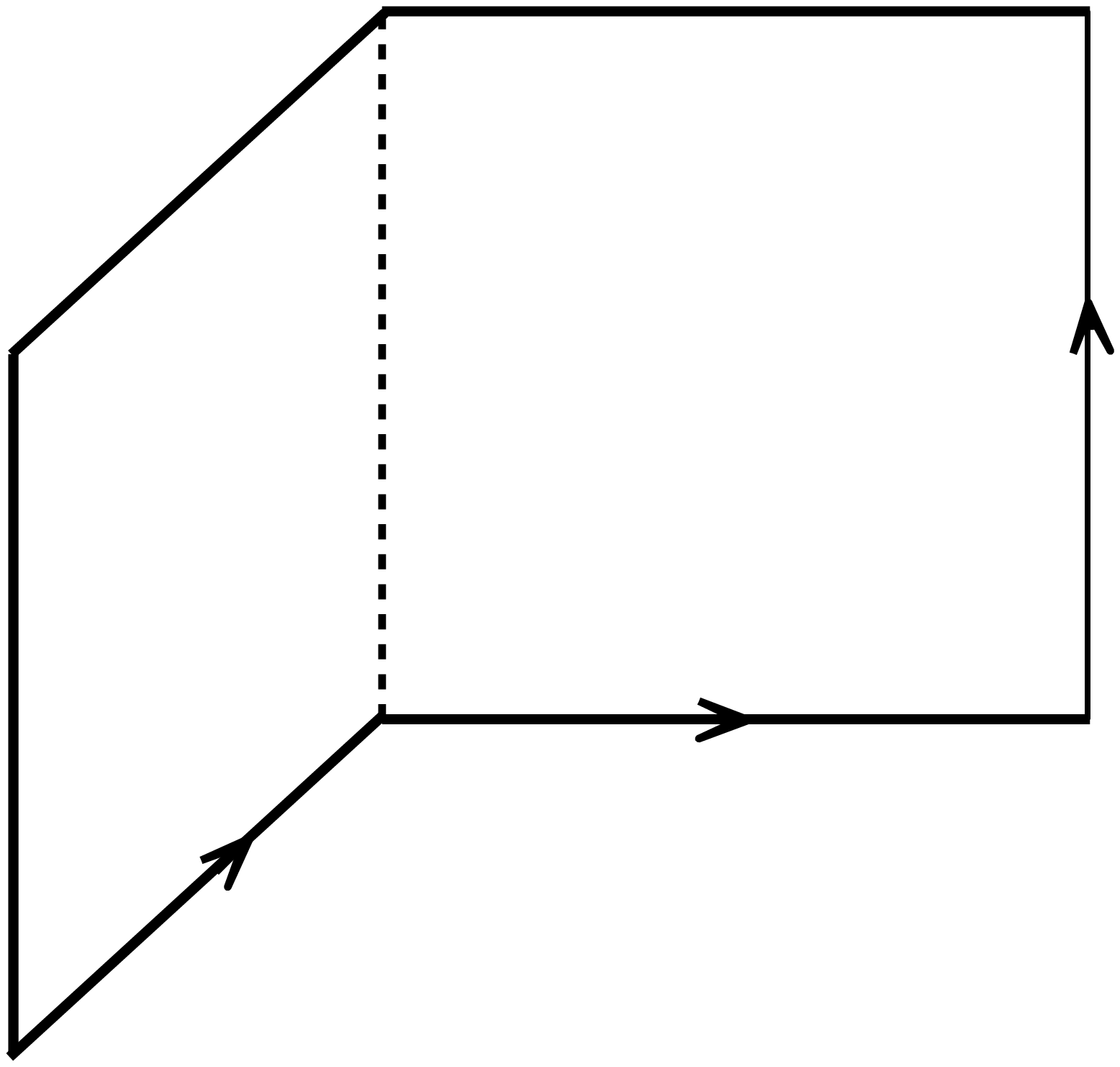}}
\hspace{-15mm}
\subfigure[]{\includegraphics[scale=0.251]{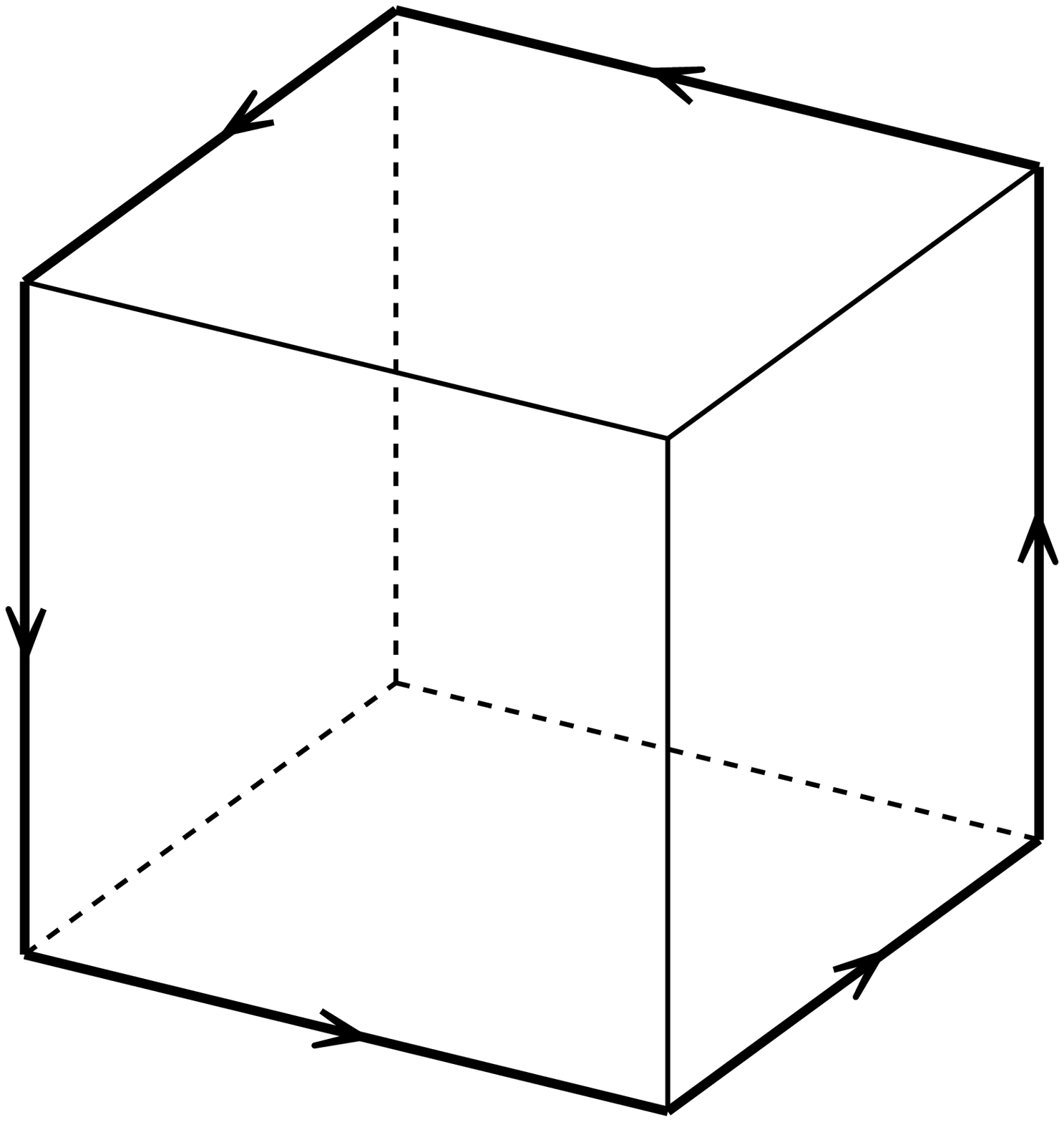}}
\hspace{-15mm}
\caption{Four types of Wilson loops in the action.}
\label{fig1}
\end{figure}
We take a four-dimensional hyper-cubic lattice and denote the sites of the lattice by an integer-valued four vector $n$. The link variables of the SU(N) 
lattice gauge theory are $N \times N$ unitary matrices and denoted by $U_\mu(n)$ for the oriented link from $n$ to $n+\hat{\mu}$. For the oriented link from 
$n+\hat{\mu}$ to $n$ we assign $U^{-1}_\mu(n)$.

There are infinitely many choices for lattice actions which give the same naive continuum limit. For example, we adopt four coupling gauge action, 
including up to six-link loop:
\begin{eqnarray}
\label{eq.2-1}
S = \frac{1}{g^2} \{ c_0 \sum Tr\left(\text{simple plaquette loop}\right) &+& c_1 \sum Tr\left(\text{rectangle loop}\right) + c_2 \sum Tr\left(\text{chair-type loop}\right) 
\nonumber\\
&+& c_3 \sum Tr\left(\text{three-dimensional loop}\right) + \text{constant}\}.
\end{eqnarray}
Four type of loop are depicted in Fig.\ref{fig1}. In sum over loops, each oriented loop appears once.

In the week coupling expansion the link variable is parameterized as
\begin{equation}
\label{eq.2-2}
U_\mu(n) = e^{i a g A_\mu(n)}
\end{equation}
end the action is expanded in terms of the field $A_\mu(n)$. Requiring that the action (\ref{eq.2-1}) reduces to 
\begin{equation}
\label{eq.2-3}
S = -\frac{1}{4} \int{d^4x \sum_{\mu,\nu,a}\left( F^a_{\mu\nu} \right)^2}
\end{equation}
in the naive continuum limit, we obtain the normalization condition
\begin{equation}
\label{eq.2-4}
c_0 + 8c_1 + 16c_2 + 8c_3 = 1.
\end{equation}
We have to fix the gauge in perturbation calculation. We take the lattice Lorentz gauge given by
\begin{equation}
\label{eq.2-5}
S_{gf} = -a^2\sum_{n}{\sum_\mu{Tr \left( \nabla_\mu A_\mu \right(n))^2}}.
\end{equation}
Let us introduce the following Fourier transformation
\begin{equation}
\label{eq.2-6}
A_\mu(n) = \int_k{e^{ikn+ik_\mu/2}A_\mu(k)}
\end{equation}
where $\int_k$ denotes $\prod^4_{\mu=1}{\int_{-\pi}^{\pi}{\frac{dk_\mu}{2\pi}}}$. Here and hereafter we make a change of variable from $k_\mu a$ to $k_\mu$.

The free part of the action $S_0$ is given by \cite{Weisz}
\begin{equation}
\label{eq.2-7}
S_0 = -\frac{1}{2}\int{\sum_{k,\mu,\nu}{Tr \left[ q_{\mu\nu}(k)\hat{F}_{\mu\nu}(k)\hat{F}_{\mu\nu}(-k) + 2\hat{k}_\mu\hat{k}_\nu A_\mu(k)A_\nu(-k) \right]}}
\end{equation}
where
\begin{equation}
\label{eq.2-8}
\hat{k}_\mu = 2 \sin{\frac{k_\mu}{2}},
\end{equation}
\begin{equation}
\label{eq.2-9}
\hat{F}_{\mu\nu} = i\left( \hat{k}_\mu A_\nu(k) - \hat{k}_\nu A_\mu(k) \right),
\end{equation}
and
\begin{eqnarray}
\label{eq.2-10}
q_{\mu\nu}(k) &=& 1 - \left( c_1 - c_2 - c_3 \right) \left( \hat{k}^2_\mu + \hat{k}^2_\nu \right) - \left( c_2 + c_3 \right) \hat{k}^2; \quad \text{for } \mu \neq \nu,
\nonumber\\
q_{\mu\mu}(k) &=& 0; \quad \text{for all } \mu.
\end{eqnarray}
Here
\begin{equation}
\label{eq.2-11}
\hat{k}^2 = \sum^4_{\mu=1}{\hat{k}^2_\mu}.
\end{equation}
The free propagator is defined by
\begin{equation}
\label{eq.2-12}
\left< A^a_\mu(k)A^b_\nu(k') \right> = \delta_{ab}(2\pi)^4\delta^{(4)}(k+k')D_{\mu\nu}(k),
\end{equation}
and is the inverse of the free two-point function
\begin{equation}
\label{eq.2-13}
G_{\mu\nu}(k) = \hat{k}_\mu\hat{k}_\nu - \sum_\rho\left( \hat{k}_\rho\delta_{\mu\nu} - \hat{k}_\mu\delta_{\rho\nu} \right) q_{\mu\rho} \hat{k}_\rho.
\end{equation}
The explicit form of the propagator is given in the Appendix \ref{appendixA}. Note that the free part of action depends on only the combinations of $c_1-\left( c_2+c_3 \right)$ 
and $c_2 + c_3$. Therefore we have two parameters $c_1$ and $c_{23} = c_2 + c_3$ to choose in the lowest order perturbation theory, because $c_0$ is determined 
from them by the constraint (\ref{eq.2-4}).

We will make the perturbation calculation with this free action. Let us denote the expectation value in the perturbation theory as
\begin{equation}
\label{eq.2-14}
\left<O\right> = \int{DA O e^{S_0}}\Big/\int{DA e^{S_0}}.
\end{equation}
The free propagator of $F_{\mu\nu}(k)$ is defined by
\begin{equation}
\label{eq.2-15}
\left< F^a_{\mu\nu}(k)F^b_{\rho\lambda}(k') \right> = \delta_{ab}(2\pi)^4\delta^{(4)}(k+k') D_{\mu\nu,\rho\lambda}(k).
\end{equation}
For example, $D_{\mu\nu,\rho\lambda}(k)$ is given by
\begin{equation}
\label{eq.2-16}
D_{\mu\nu,\rho\lambda}(k) = \left( \hat{k}_\mu \right)^2 D_{\nu\nu} + \left( \hat{k}_\nu \right)^2 D_{\mu\mu} - 2\hat{k}_\mu\hat{k}_\nu D_{\mu\nu}.
\end{equation}

The expectation value of various Wilson loops can be written in the lowest order of perturbation theory as
\begin{equation}
\label{eq.2-17}
W(C) = 1 - g^2\frac{N^2-1}{4N}F(C)
\end{equation}
where C denotes the Wilson loop. Note that $F(C)$ is independent of $N$. The $F(C)$ for some particular Wilson loops are given as follows:
\begin{subequations}
\label{eq.2-18}
\begin{equation}
\label{eq.2-18a}
F(I\times J) = \int_k{D_{12,12} \left( \frac{\sin{\frac{1}{2}Ik_1}}{\sin{\frac{1}{2}k_1}} \frac{\sin{\frac{1}{2}Jk_2}}{\sin{\frac{1}{2}k_2}} \right)^2}
\end{equation}
\begin{equation}
\label{eq.2-18b}
F(\text{chair}) = \int_k{D_{12,12} \left( \frac{3}{2} + \frac{1}{2}\cos{k_3} \right)}
\end{equation}
\begin{equation}
\label{eq.2-18c}
F(\text{3-dim.}) = \int_k{D_{12,12} \left( \frac{3}{2} + \frac{3}{2}\cos{k_3} \right)}
\end{equation}
\begin{equation}
\label{eq.2-18d}
F(\text{4-dim.}) = \int_k{D_{12,12} \left[ 6 - 2\left(1-\cos{k_3}\right)\left(2+\cos{k_4}\right) \right]}
\end{equation}
\end{subequations}
Here $F(I\times J)$ is for $I\times J$ rectangular loop, $F(\text{chair})$ is for the loop depicted in Fig.\ref{fig1}.c, $F(\text{3-dim.})$ for the loop in 
Fig.\ref{fig1}.d. The 4-dimensional loop in eq.(\ref{eq.2-18d}) may be described by the loop:
\begin{eqnarray}
\label{eq.4-dim}
n \;&\rightarrow&\; n+\hat{1} \;\rightarrow\; n+\hat{1}+\hat{2} \;\rightarrow\; n+\hat{1}+\hat{2}+\hat{3} \;\rightarrow\; n+\hat{1}+\hat{2}+\hat{3}+\hat{4}
\nonumber\\
\;&\rightarrow&\; n+\hat{2}+\hat{3}+\hat{4} \;\rightarrow\; n+\hat{3}+\hat{4} \;\rightarrow\; n+\hat{4} \;\rightarrow\; n.
\nonumber
\end{eqnarray}

\section{Block spin transformation}
\label{section3}
\begin{figure}[t]
\vspace{-30mm}
{\includegraphics[scale=0.6]{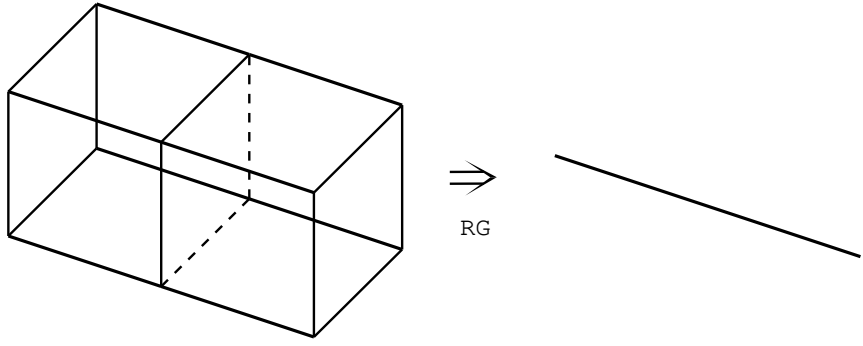}}
\vspace{-50mm}
\caption{Schematic diagram for block spin transformation (3-dimensional projection).}
\label{fig2}
\end{figure}

First note a difference between 2d O(N) sigma model and 4d SU(N) gauge model in that the variables are site-variables for the former, while they are link-variables 
for the later. Therefore we propose to make blocks of link variables in 4d SU(N) gauge models, rather than blocks of sites. See Fig.\ref{fig2}.

Secondly we are interested in the limit $g \rightarrow g_c (=0)$ and in the physical quantities for which perturbation theory is applicable. As discussed in the 
first section of [I], the correlation length is large near the critical point $g_c$ and diverges at the critical point as a result of cooperative behavior of the 
system. Within the correlation length the properties of the system do not change qualitatively. Therefore, if we make the weak coupling expansion, the variables 
$A_\mu(n)$ do not change rapidly within the correlation length.

Thus we propose to define a block transformation by
\begin{equation}
\label{eq.3-1}
A^{(1)}_\mu(n) = \frac{1}{8}\left[ A_\mu(2n) +\sum^4_{\nu=1}{A_\mu(2n+\hat{\nu})} +\sum_{\nu \neq \rho}{A_\mu(2n+\hat{\nu}+\hat{\rho})}
 +\sum_{\nu \neq \rho \neq \lambda \neq \nu}{A_\mu(2n+\hat{\nu}+\hat{\rho}+\hat{\lambda})} \right]
\end{equation}
and
\begin{equation}
\label{eq.3-2}
U^{(1)}(n) = \exp\left( i g a A^{(1)}_\mu(n) \right).
\end{equation}
Note that the normalization constant is $1/8$ rather than $1/16$, because the length of the link in the new system is $2a$. Except for the normalization, 
eq.(\ref{eq.3-1}) is similar to eq.(3.1) in (I). Therefore let us write eq.(3.1) simply as
\setcounter {equation} {0}
\begin{subequations}
\label{eq.3-1a}
\begin{equation}
A^{(1)}_\mu(n') = \frac{1}{8}\sum_{n \in n'} A_\mu(n),
\end{equation}
\end{subequations}
\setcounter {equation} {2}
and let us call this transformation block spin transformation, although the variable is not spin-variable. The block spin transformation is applied repeatedly. 
A block variable $A^{(I)}_\mu(n)$ after $I-th$ iteration is defined as
\begin{equation}
\label{eq.3-3}
A^{(I)}_\mu(n') = \frac{1}{8}\sum_{n \in n'} A^{(I-1)}_\mu(n)
\end{equation}
and
\begin{equation}
\label{eq.3-4}
U^{(I)}(n) = \exp\left( i g a A^{(I)}_\mu(n) \right).
\end{equation}
The effective action after the block spin transformation is defined through the relation
\begin{equation}
\label{eq.3-5}
e^{S^{(I)}\left(A^{(I)}\right)} = \int_{\{A^{(I-1)}\}}{K\left(A^{(I-1)},A^{(I)}\right)e^{S^{(I-1)}\left(A^{(I-1)}\right)}}
\end{equation}
where
\begin{equation}
\label{eq.3-6}
K\left(A^{(I-1)},A^{(I)}\right) = \prod_\mu \prod_{n'}{\delta\left(A_\mu(n') - \frac{1}{8}\sum_{n \in n'}{A_\mu(n)}\right)}.
\end{equation}
The renormalization group $\tau$ is defined as
\begin{equation}
\label{eq.3-7}
\tau\left(S^{(I-1)}\right) = S^{(I)}.
\end{equation}
The $S^{(0)}$ is the original action and $A^{(0)}_\mu(n)$ is the original variable $A_\mu(n)$.
This transformation is slightly different from that proposed by Wilson \cite{Wilson1980}, although it is the same in spirit. This difference becomes, however, 
crucial when ones consider the limit $I \rightarrow \infty$, as will be shown in the next section.

Let us calculate the expectation value of various Wilson loops for the $I-th$ block link variables in the perturbation theory. Let us denote them as
\begin{equation}
\label{eq.3-8}
W{(I)}(C) = 1 - g^2\frac{N^2-1}{4N}F^{(I)}(C)
\end{equation}
following eq.(\ref{eq.2-17}). The expectation value is taken with the action $S^{(I)}\left(A^{(I)}\right)$. As is in 2d O(N) sigma model, it is easy to calculate 
$F^{(I)}(C)$ in terms of the original variables in the perturbation theory. We obtain
\begin{subequations}
\label{eq.3-9}
\begin{equation}
\label{eq.3-9a}
F^{(I)}(I\times J) = \int{D^{(I)}_{12,12}\left( \frac{\sin{\frac{1}{2}I k^{(I)}_1}}{\sin{\frac{1}{2}k^{(I)}_1}}
\frac{\sin{\frac{1}{2}J k^{(I)}_2}}{\sin{\frac{1}{2}k^{(I)}_2}} \right)^2 H^{(I)}(k)},
\end{equation}
\begin{equation}
\label{eq.3-9b}
F^{(I)}(\text{chair}) = \int{D^{(I)}_{12,12}\left( \frac{3}{2}+\frac{1}{2}\cos{k^{(I)}_3} \right) H^{(I)}(k)},
\end{equation}
\begin{equation}
\label{eq.3-9c}
F^{(I)}(\text{3-dim}) = \int{D^{(I)}_{12,12}\left( \frac{3}{2}+\frac{3}{2}\cos{k^{(I)}_3} \right) H^{(I)}(k)},
\end{equation}
\begin{equation}
\label{eq.3-9d}
F^{(I)}(\text{4-dim}) = \int{D^{(I)}_{12,12}\left[ 6 - 2 \left(1-\cos{k^{(I)}_3}\right)\left(2+\cos{k^{(I)}_4}\right) \right] H^{(I)}(k)}.
\end{equation}
\end{subequations}
Here
\begin{equation}
\label{eq.3-10}
k^{(I)}_\mu = 2^Ik_\mu \quad (\mu=1 \sim 4)
\end{equation}
\begin{equation}
\label{eq.3-11}
H^{(I)}(k) = \prod^{I-1}_{M=0}{\frac{1}{4}\prod^4_{\mu=1}\left(1+\cos 2^M k_\mu\right)}
\end{equation}
and
\begin{eqnarray}
\label{eq.3-12}
D^{(I)}_{\mu\nu,\mu\nu}(k) &=& \left(\hat{k}^{(I)}_\mu\right)^2 D_{\nu\nu}(k) + \left(\hat{k}^{(I)}_\nu\right)^2 D_{\mu\mu}(k)
\nonumber\\
&-& 2 \hat{k}^{(I)}_\mu \hat{k}^{(I)}_\nu \cos \left(\left(2^{I-1}-\frac{1}{2}\right)k_\mu\right) \cos \left(\left(2^{I-1}-\frac{1}{2}\right)k_\nu\right)
D_{\mu\nu}(k)
\end{eqnarray}
where
\begin{equation}
\label{eq.3-13}
\hat{k}^{(I)}_\mu = 2 \sin {\frac{k^{(I)}_\mu}{2}}.
\end{equation}
The derivation  of eq.(\ref{eq.3-9}) is given in the Appendix \ref{appendixB}.

The gauge for the block variables $A^{(I)}_\mu(n)\quad(I\ge1)$ is not the lattice Lorentz gauge, in general. This is not the problem to calculate $F^{(I)}(C)$, 
as is shown above.

\section{Expectation value of various Wilson loops on the renormalized trajectory}
\label{section4}
In the previous section we have obtained the expectation value of various Wilson loops for block link variables after $I-th$ iteration. In the limit 
$I \rightarrow \infty$, the effective action $S^{(I)}$ should approach the renormalized trajectory. Therefore the expectation value of the Wilson loop 
also should approach to that on the renormalized trajectory. Let us denote
\begin{equation}
\label{eq.4-1}
\lim_{I \rightarrow \infty }{W^{(I)}(C)} = W^{(\infty)}(C) = 1 - g^2\frac{N^2-1}{4N}F^{(\infty)}(C)
\end{equation}
In this section we will derive $F^{(\infty)}(C)$.

First note
\begin{equation}
\label{eq.4-2}
\lim_{n \rightarrow \infty}\int{e^{i k n}D_{\mu\nu}(k)} = \frac{1}{4\pi^2}\frac{1}{n^2}\delta_{\mu\nu} + O\left(\frac{1}{n^4}\right)
\end{equation}
The first term in the r.h.s. is the dominant contribution to the expectation value in the limit $I \rightarrow \infty$. As in the section 4 of [I], we are able 
to prove that the non-leading terms which come from the second term of the r.h.s. of eq.(\ref{eq.4-2}) do not contribute to the expectation value in the limit 
$I \rightarrow \infty$. Hence we neglect them. Therefore we have to consider only $D_{\mu\nu}(k)$ with $\mu=\nu$.\\
Secondly let us remind that
\begin{equation}
\label{eq.4-3}
A^{(I)}(l,m,n,\tau) = \frac{1}{8^I}\sum^{2^I-1}_{\substack{l_1,m_1\\n_1,\tau_1}}{A^{(0)}\left(2^I l + l_1, 2^I m + m_1, 2^I n + n_1, 2^I \tau + \tau_1\right)}.
\end{equation}
Now let us consider, for example, $F^{(I)}(1\times 1)$. From eqs.(\ref{eq.4-2}) and (\ref{eq.4-3}) we have
\begin{eqnarray}
\label{eq.4-4}
F^{(I)}(1\times 1) \sim 4\left(\frac{1}{8^I}\right)\sum_{\substack{l_1,m_1,l_2,m_2\\n_1,\tau_1,n_2,\tau_2}} 
\Big[ &F&\left(l_1-l_2,m_1-m_2,n_1-n_2,\tau_1-\tau_2\right) \nonumber\\
- &F&\left(2^I + l_1-l_2,m_1-m_2,n_1-n_2,\tau_1-\tau_2\right) \Big]
\end{eqnarray}
where
\begin{equation}
\label{eq.4-5}
F(l,m,n,\tau) = \frac{1}{4\pi^2}\frac{1}{l^2+m^2+n^2+\tau^2}.
\end{equation}
The approximate equality in eq.(\ref{eq.4-4}) means that only the leading terms are taken. Equation (\ref{eq.4-4}) may be written as
\begin{eqnarray}
\label{eq.4-6}
F^{(I)}(1\times 1) \sim 4\left(\frac{1}{16^I}\right)^2\frac{1}{4\pi^2} \sum\sum \Bigg[
\frac{1}{\left(\frac{l_1-l_2}{2^I}\right)^2 + \left(\frac{m_1-m_2}{2^I}\right)^2 + \left(\frac{n_1-n_2}{2^I}\right)^2 + \left(\frac{\tau_1-\tau_2}{2^I}\right)^2}
\nonumber\\
- \frac{1}{\left(1+\frac{l_1-l_2}{2^I}\right)^2 + \left(\frac{m_1-m_2}{2^I}\right)^2 + \left(\frac{n_1-n_2}{2^I}\right)^2 
+ \left(\frac{\tau_1-\tau_2}{2^I}\right)^2}\Bigg]
\end{eqnarray}
Note that the factor $(1/2)^{2I}$ are multiplied for the denominator and the coefficient. Putting $1/2^I = \epsilon$ as in [I], we obtain finally 
\begin{eqnarray}
\label{eq.4-7}
F^{(\infty)}(1\times 1) &=& 4\frac{1}{4\pi^2}\int dx_1dy_1dz_1d\tau_1 \int  dx_2dy_2dz_2d\tau_2\nonumber\\
&&\Bigg(\frac{1}{\left(x_1-x_2\right)^2 + \left(y_1-y_2\right)^2 + \left(z_1-z_2\right)^2 + \left(\tau_1-\tau_2\right)^2} \nonumber\\
&&- \frac{1}{\left(1+x_1-x_2\right)^2 + \left(y_1-y_2\right)^2 + \left(z_1-z_2\right)^2 + \left(\tau_1-\tau_2\right)^2}\Bigg)
\end{eqnarray}
Introducing the function $\tilde{\tilde{F}}$ defined by
\begin{equation}
\label{eq.4-8}
\tilde{\tilde{F}}(n_1,n_2,n_3,n_4) =\frac{1}{4\pi^2}\sum_{\substack{\epsilon_1,\epsilon_2,\epsilon_3,\epsilon_4\\ \pm 1}}
\frac{1}{ \left(n_1-\epsilon_1x\right)^2 + \left(n_2-\epsilon_2y\right)^2 + \left(n_3-\epsilon_3z\right)^2 + \left(n_4-\epsilon_4\tau\right)^2 }
\end{equation}
and setting $x_1-x_2 = x$ and so on, we obtain
\begin{equation}
\label{eq.4-9}
F^{(\infty)}(1\times 1) = 4 \iiiint\limits^1_0dxdydzd\tau(1-x)(1-y)(1-z)(1-\tau)\times\left(\tilde{\tilde{F}}(0,0,0,0) - \tilde{\tilde{F}}(1,0,0,0)\right).
\end{equation}
We are able to obtain other $F^{(\infty)}(C)$ similarly.

Let us introduce the symbol
\begin{equation}
\label{eq.4-10}
\sqiint = \iiiint\limits^1_0(1-x)(1-y)(1-z)(1-\tau)dxdydzd\tau
\end{equation}
for notational simplicity. Then we have, for example,
\begin{eqnarray}
\label{eq.4-11}
F^{(\infty)}(1\times 1) &=& 4\sqiint \left(\tilde{\tilde{F}}(0,0,0,0) - \tilde{\tilde{F}}(1,0,0,0)\right) \nonumber\\
F^{(\infty)}(1\times 2) &=& \sqiint \left[6\tilde{\tilde{F}}(0,0,0,0) - 4\tilde{\tilde{F}}(1,1,0,0) -\tilde{\tilde{F}}(2,0,0,0)\right] \nonumber\\
F^{(\infty)}(2\times 2) &=& \sqiint 8\left[\tilde{\tilde{F}}(0,0,0,0) + \tilde{\tilde{F}}(1,0,0,0) - \tilde{\tilde{F}}(2,0,0,0) - \tilde{\tilde{F}}(2,1,0,0)\right] 
\nonumber\\
F^{(\infty)}(\text{chair}) &=& \sqiint \left[6\tilde{\tilde{F}}(0,0,0,0) - 4\tilde{\tilde{F}}(1,0,0,0) -2\tilde{\tilde{F}}(1,1,0,0)\right] \nonumber\\
F^{(\infty)}(\text{3-dim}) &=& \sqiint 6\left[\tilde{\tilde{F}}(0,0,0,0) - \tilde{\tilde{F}}(1,1,0,0)\right] \nonumber\\
F^{(\infty)}(\text{4-dim}) &=& \sqiint 6\left[\tilde{\tilde{F}}(0,0,0,0) - \tilde{\tilde{F}}(1,1,1,0)\right]
\end{eqnarray}
It should be noted that it is non-trivial that $F^{(I)}(C)$ approach finite constant value in the limit $I \rightarrow \infty$. It depends on the definition 
of renormalization group. If we had taken the renormalization group proposed in the ref.\cite{Wilson1980}, we would obtain that $F^{(\infty)}(C)$ are 
trivially zero: Because the length of the new link after the renormalization does not change from that of the original link, we have the sum over only $l_1$, 
$m_1$, and $n_1$ in the equation corresponding to eq.(\ref{eq.4-3}) and consequently we have an extra factor $(1/2^I)^2$ which vanishes in the limit 
$I \rightarrow \infty$, in the equation corresponding to eq.(\ref{eq.4-7}).

\section{Block spin transformation for various actions}
\label{section5}
\begin{figure}[t]
\centering
\vspace{-12mm}
\subfigure[]{\includegraphics[scale=0.6]{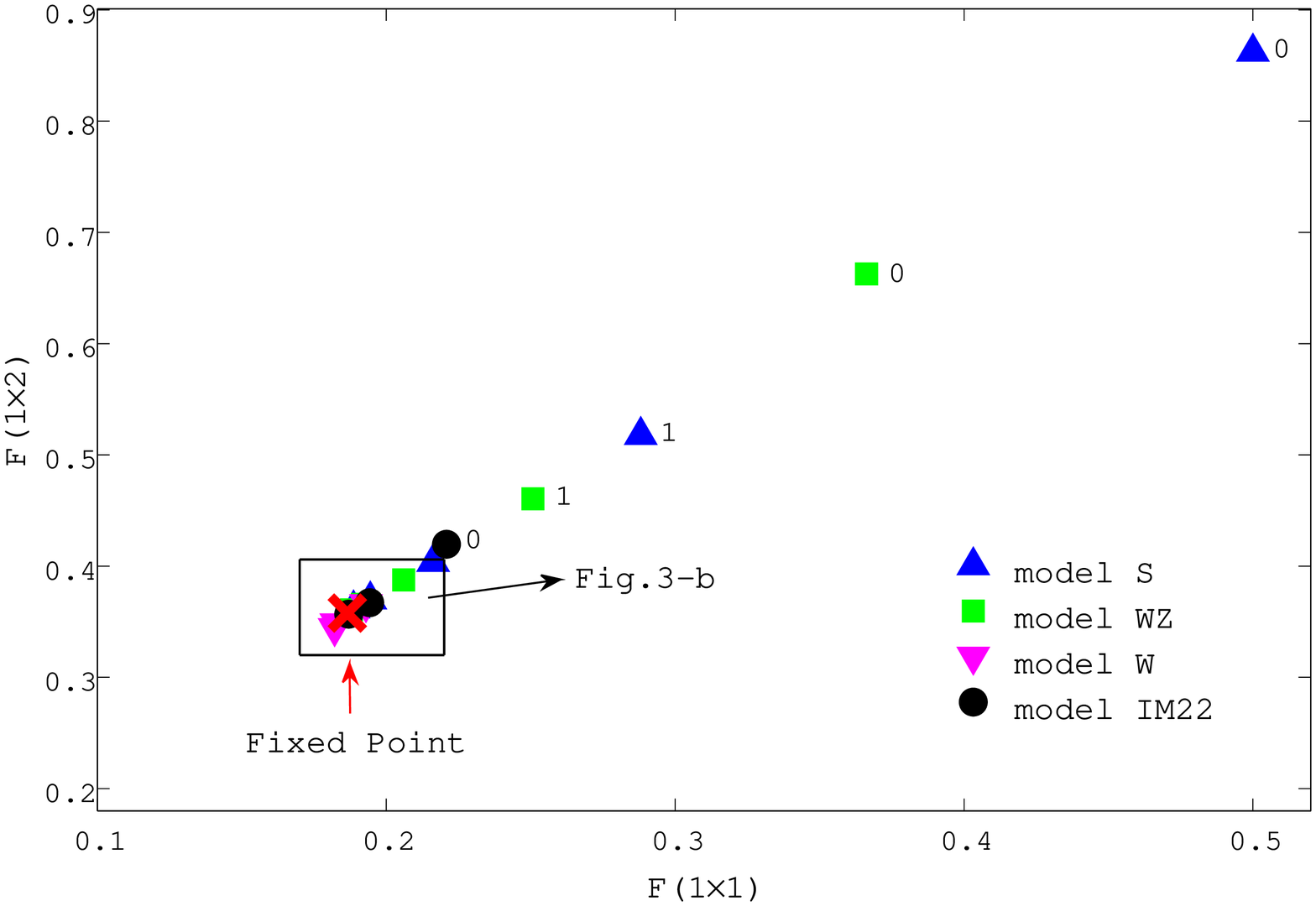}}
\vspace{-2mm}
\subfigure[]{\includegraphics[scale=0.6]{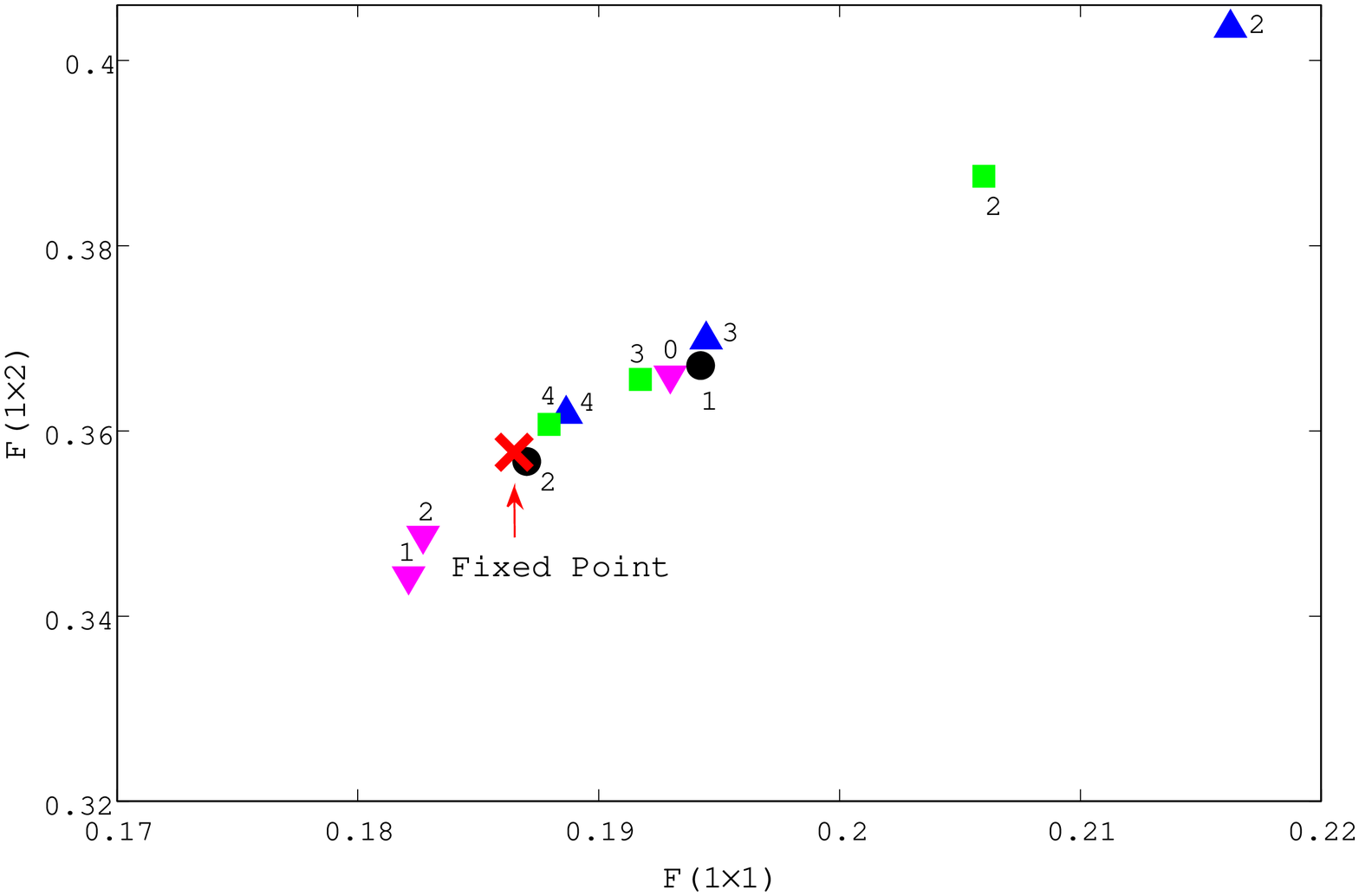}}
\caption{Block spin renormalization group flows for $F^{(I)}(1\times1)$ and $F^{(I)}(1\times2)$ for various models. The numbers from 0 to 4 correspond to $I$ of 
$F^{(I)}(n\times m)$.}
\label{fig3}
\end{figure}

Now we are ready to calculate the function $F^{(I)}(C)$ of the block variables for any parameter $c_1$ and $c_{23} \; (=c_2+c_3)$, 
using eqs.(\ref{eq.3-9})$\sim$(\ref{eq.3-13}). We first list some of them for the standard model $\left(c_1=c_2=c_3=0\right)$ together with $F^{(\infty)}(C)$ obtained 
from eqs.(\ref{eq.4-10}) and (\ref{eq.4-11}). We clearly see from the table \ref{table1} that the functions gradually approach the asymptotic values. We also list some of 
$F^{(I)}(C)$ in the table \ref{table1} for $c_1=-0.252, c_2=0,c_3=-0.17$ (model W) which has been chosen by Wilson in ref.\cite{Wilson1980}, for $c_1=-1/12, c_2=c_3=0$ 
(model WZ) which has been chosen by Weisz \cite{Weisz} in Symanzik approach \cite{Symanzik}, as well as for $c_1=-0.331, c_2=c_3=0$ (model IM11) and for 
$c_1=-0.27, c_{23}=-0.04$ (model IM22) which will be chosen by a certain criterion below. In Fig.\ref{fig3} we depict block spin transformation flows for the function 
$F^{(I)}(C)$.

We  see from the Figure and the Table that the behavior of the convergence for the function to the fixed point crucially on the parameter $c_1$ and $c_{23}$.

\begin{table}[t]
\caption{The values of $F^{(I)}(C)$ for various models together with $F^{(\infty)}(C)$.}
\label{table1}
{Table 1-a \quad\quad Model S ($c_1=c_2=c_3=0$)}
\centering
\begin{tabular}{| c | c | c | c | c | c | c |}
\hline
& F($1\times1$) & F($1\times2$) & F($2\times2$) & F(chair)&F(3-dim.) & F(4-dim.)\\
\hline
$F^{(0)}$      & 0.50000 & 0.86225 & 1.36931 & 0.78444 & 0.85331 & 1.17759\\
$F^{(1)}$      & 0.28810 & 0.51765 & 0.87978 & 0.45667 & 0.50569 & 0.70854\\
$F^{(2)}$      & 0.21623 & 0.40352 & 0.72087 & 0.34655 & 0.39094 & 0.55476\\
$F^{(3)}$      & 0.19446 & 0.36985 & 0.67515 & 0.31350 & 0.35711 & 0.50956\\
$F^{(4)}$      & 0.18865 & 0.36183 & 0.66877 & 0.30465 & 0.34801 & 0.49680\\
\hline
$F^{(\infty)}$ & 0.18649 & 0.35770 & 0.65875 & 0.30146 & 0.34493 & 0.49331\\
\hline
\end{tabular}
\begin{tabular}{c}
\\{Table 1-b \quad\quad Model W ($c_1=-0.252, c_2=0, c_3=-0.17$)}
\end{tabular}
\centering
\begin{tabular}{| c | c | c | c | c | c | c |}
\hline
& F($1\times1$) & F($1\times2$) & F($2\times2$) & F(chair)&F(3-dim.) & F(4-dim.)\\
\hline
$F^{(0)}$      & 0.19297 & 0.36601 & 0.64859 & 0.31059 & 0.35287 & 0.50047\\
$F^{(1)}$      & 0.18211 & 0.34432 & 0.62927 & 0.29439 & 0.33685 & 0.48168\\
$F^{(2)}$      & 0.18271 & 0.34871 & 0.64290 & 0.29573 & 0.33906 & 0.48554\\
\hline
$F^{(\infty)}$ & 0.18649 & 0.35770 & 0.65875 & 0.30146 & 0.34493 & 0.49331\\
\hline
\end{tabular}
\begin{tabular}{c}
\\{Table 1-c \quad\quad Model WZ ($c_1=-1/12, c_2=c_3=0$)}
\end{tabular}
\centering
\begin{tabular}{| c | c | c | c | c | c | c |}
\hline
& F($1\times1$) & F($1\times2$) & F($2\times2$) & F(chair)&F(3-dim.) & F(4-dim.)\\
\hline
$F^{(0)}$      & 0.36626 & 0.66263 & 1.09814 & 0.57705 & 0.63235 & 0.87786\\
$F^{(1)}$      & 0.25076 & 0.46043 & 0.80091 & 0.39900 & 0.44470 & 0.62629\\
$F^{(2)}$      & 0.20599 & 0.38751 & 0.69875 & 0.33087 & 0.37464 & 0.53292\\
$F^{(3)}$      & 0.19173 & 0.36556 & 0.66924 & 0.30934 & 0.35285 & 0.50387\\
$F^{(4)}$      & 0.18794 & 0.36072 & 0.66724 & 0.30358 & 0.34691 & 0.49534\\
\hline
$F^{(\infty)}$ & 0.18649 & 0.35770 & 0.65875 & 0.30146 & 0.34493 & 0.49331\\
\hline
\end{tabular}
\begin{tabular}{c}
\\{Table 1-d \quad\quad Model IM11 ($c_1=-0.331, c_2=c_3=0$)}
\end{tabular}
\centering
\begin{tabular}{| c | c | c | c | c | c | c |}
\hline
& F($1\times1$) & F($1\times2$) & F($2\times2$) & F(chair)&F(3-dim.) & F(4-dim.)\\
\hline
$F^{(0)}$      & 0.21027 & 0.40340 & 0.71109 & 0.33352 & 0.36977 & 0.51874\\
$F^{(1)}$      & 0.18826 & 0.35756 & 0.65039 & 0.30184 & 0.34074 & 0.48500\\
$F^{(2)}$      & 0.18431 & 0.35247 & 0.64945 & 0.29743 & 0.33935 & 0.48528\\
\hline
$F^{(\infty)}$ & 0.18649 & 0.35770 & 0.65875 & 0.30146 & 0.34493 & 0.49331\\
\hline
\end{tabular}
\end{table}

\setcounter {table} {0}
\begin{table}[t]
\caption{(Continue)}
{Table 1-e \quad\quad Model IM22 ($c_1=-0.27, c_{23}=-0.04$)}
\centering
\begin{tabular}{| c | c | c | c | c | c | c |}
\hline
& F($1\times1$) & F($1\times2$) & F($2\times2$) & F(chair)&F(3-dim.) & F(4-dim.)\\
\hline
$F^{(0)}$      & 0.22081 & 0.41968 & 0.73586 & 0.35136 & 0.39167 & 0.55003\\
$F^{(1)}$      & 0.19423 & 0.36707 & 0.66472 & 0.31177 & 0.35263 & 0.50167\\
$F^{(2)}$      & 0.18701 & 0.35671 & 0.65556 & 0.30181 & 0.34439 & 0.49220\\
\hline
$F^{(\infty)}$ & 0.18649 & 0.35770 & 0.65875 & 0.30146 & 0.34493 & 0.49331\\
\hline
\end{tabular}
\end{table}

\section{Improved lattice action}
\label{section6}
According to our strategy described in detail in [I], let us choose an action which is located near the renormalized trajectory. First let us note that there is 
no difference between interaction, for example, $Tr\left(U_1 U_2 U_3 U_4\right)Tr\left(U_4^{-1} U_5 U_6 U_7\right)$ and $Tr\left(U_1 U_2 U_3 U_5 U_6 U_7\right)$ 
in the lowest order perturbation theory. Therefore we assume that the action is of the form
\begin{equation}
\label{eq.6-1}
S^{(\infty)} = \sum_C{K_C W(C)}
\end{equation}
where $W(C)$ is the Wilson loop for the contour $C$. Practically we have to truncate the sum in eq.(\ref{eq.6-1}). Let us restrict actions to those given by 
eq.(\ref{eq.2-1}).

We define a distance from an action to the renormalized trajectory by
\begin{equation}
\label{eq.6-2}
R^{(I)} = \sqrt{\sum_C{\left(\frac{F^{(I)}(C)-F^{(\infty)}(C)}{F^{(\infty)}(C)}\right)^2\Big/M}}
\end{equation}
where $M$ is the number of terms in the sum over the contour $C$. We restrict the contour in the sum to those up to six-link length.

Plotting $R^{(1)}$ and $R^{(2)}$ defined above in the two-dimensional space spanned by $c_1$ and $c_{23}$, we find that there is a one-dimensional very narrow deep 
valley where $R^{(1)}$ and $R^{(2)}$ are very small, respectively. For $R^{(1)}$, the one-dimensional line is parameterized by the equation
\begin{equation}
\label{eq.6-3}
1.531c_1 + c_{23} = -0.5067
\end{equation}
and for $R^{(2)}$
\begin{equation}
\label{eq.6-4}
1.746c_1 + c_{23} = -0.5116.
\end{equation}

Along the line defined by eq.(\ref{eq.6-3}), $R^{(1)}$ is less than $0.01$ and the variation of $R^{(1)}$ is not so rapid as far as $0\gtrsim c_{23} \gtrsim -0.2$. 
The minimum of $R^{(1)}$ is about $0.00545$ at $c_1=-0.27$ and $c_{23}=-0.09333$ (to be referred as point IM12). When we put $c_{23} = 0$, the minimum of $R^{(1)}$ 
is about $0.00773$ at $c_1=-0.331$ (point IM11). Note that the values of $R^{(1)}$ do not differ so much between the two points.

Along the line defined by eq.(\ref{eq.6-4}), $R^{(2)}$ is less than 0.0025 and the variation of $R^{(2)}$ is not so rapid as far as $0\gtrsim c_{23} \gtrsim -0.2$. 
The minimum of $R^{(2)}$ is about $0.00219$ at $c_1=-0.27$ and $c_{23}=-0.04$ (point IM22). When we put $c_{23} = 0$, the minimum is about 0.00242 at $c_1=-0.293$ 
(point IM21).

If we could perform MC simulations with parameters $c_1$ and $c_2$ (or $c_3$) on a very large lattice, the action IM22 could be the best action among those 
considered up to here. However, it makes a large difference for computer time in MC simulations whether we include the $c_2$ (or $c_3$) term or not. On the other 
hand there is no large difference between action IM22 and action IM21, or between action IM12 and action IM11 from the view point of renormalization group. 
Therefore it is practically better to choose action IM21 or action IM11.

There is also a limitation on the size of the lattice where we make MC simulations. For example, we will measure the string tension \cite{Iwasaki-Yoshie} on a 
$8^4$ lattice for SU(3) lattice gauge theory. In this case the string tension is mainly determined by the Creutz ratio \cite{Creutz1980} $X(3,3)$ (and $X(4,4)$). 
The $X(3,3)$ is determined from $W(3\times 3), W(2\times 3)$ and $W(2\times 2)$. On the other hand, $F^{(1)}(1\times 2)$, for example, contains the contributions 
from the propagator $D_{\mu\nu}(m,n)$ up to $m=3$ and $n=5$, while $F^{(2)}(1\times 2)$ contains those up to $m=7$ and $n=11$. Thus to obtain the Wilson loops 
such as $W(3\times 3)$ or $W(4\times 4)$, the criterion based on $R^{(1)}$ is better than on $R^{(2)}$. From this consideration and an analysis of instantons on 
the lattice (see section \ref{section8}), we choose the model IM11 as an improved action (model IM). When we make MC simulations on a lattice, e. g., $16^4$, we may choose 
the model IM21 as an improved action. Anyway the difference concerning $R^{(I)}$ between models IM11 and IM21 is not large.

In ref.\cite{Iwasaki-Sakai-Yoshie}, we have measured the string tension on a $8^4$ lattice for the SU(2) lattice gauge theory with the action $c_1=-0.3371$, 
$c_2=c_3=0$ (model R3). This action has been chosen from the consideration of the scale parameter and an analysis of instantons on the lattice. For this action 
$R^{(1)} = 0.0097$. Compare this value with $R^{(1)} = 0.027$ for model W, $R^{(1)} = 0.312$ for model WZ and $R^{(1)} = 0.495$ for model S. Thus the model R3 
is also an improved action compared with models WZ and S: As far as $c_1\sim -0.3$ (with $c_{23}=0$) for which $R^{(1)} \lesssim 0.01$, any action can be taken 
as an improved action.

In ref.\cite{Iwasaki-Sakai-Yoshie} we have chosen $\ln\: \Lambda_L/\Lambda^W_L$ as a parameter for the distance between an action and action W. It seems that a line 
where $R^{(2)}$, for example, is constant corresponds approximately to a line where the scale parameter $\Lambda$ is constant, as noted in [I] for 2d O(N) sigma 
models. Thus the criterion chosen in ref.\cite{Iwasaki-Sakai-Yoshie} for the distance is reasonable even from the view point of the renormalization group.

Note that model WZ is not an improved action due to our criteria: The value $R^{(1)} = 0.312$ implies that model WZ is far from the renormalization trajectory, 
and the life time of instantons on the lattice is short. On the other hand, model W is an improved action; although $R^{(1)}$ is slightly larger than that of 
model IM11 or that of model R3, it is of the same order. 

\section{Implications for MC calculations}
\label{section7}
As far we have already discussed the general feature of the renormalization group and our strategy in section 7 of [I], we do not repeat them here. According to 
our strategy, let us set an upper limit $\Delta$ for the relative difference between $F^{(I)}(C)$ and $F^{(\infty)}(C)$ as $\Delta = 0.01$. Then for model IM we 
have $J=1$ as the minimum number of iteration for which $R^{(J)} \lesssim \Delta$ is satisfied. On the other hand for model S we have $J=4$.

This implies that the Wilson loops $W(I\times J)$ agree with $W^{(\infty)}(I\times J)$ within relative difference $0.01$ for $I,J \ge 2$ in the case of model IM, 
according to our discussion in section 7 of [I]. This further implies that the Creutz ratio $X(I,J)$ take their asymptotic value on the renormalized trajectory 
for $I,J\ge 3$.

\begin{figure}[t]
\centering
\vspace{-20mm}
\includegraphics[scale=0.6]{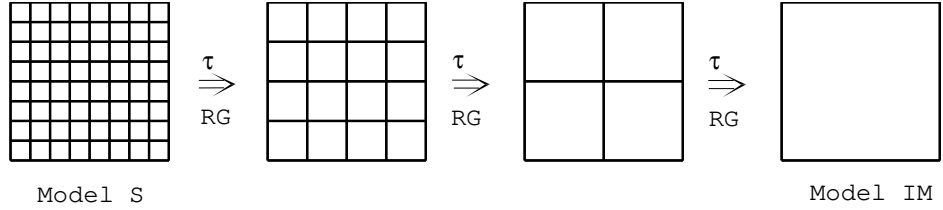}
\vspace{-25mm}
\caption{Schematic diagram of  the renormalization group from model S to model IM. The model IM in the Figure does not exactly correspond to the model IM defined 
in the text. However, they are approximately equivalent to each other, because $R^{(1)}$ for both models are of the same other.}
\label{fig4}
\end{figure}

On the other hand, the Wilson loop $W(I\times J)$ for model IM corresponds to the Wilson loop $W\left(2^3I\times 2^3J\right)$ for model S. Thus it is only expected 
that the Creutz ratios $X(I,J)$ take their asymptotic values on the renormalization trajectory, for $I,J \ge 16$ with precision $0.01$ in the case of model S. 
(See Fig.\ref{fig4}) This is the reason why we call IM the improved action.

\section{Instantons on the lattice and the renormalized trajectory}
\label{section8}
\begin{figure}[t]
\centering
\vspace{-10mm}
\includegraphics[scale=0.6]{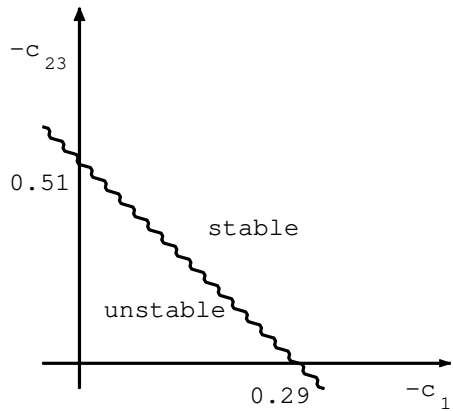}
\vspace{-5mm}
\caption{Stability of instantons on the lattice vs. $c_1$ and $c_{23}$.}
\label{fig5}
\end{figure}

The general discussion on instantons on the lattice given in section 10 of [I] can be also applied for 4d SU(N) gauge models. Thus we conclude also for 4d SU(N) 
gauge models that the renormalized trajectory is located at the boundary which devides the parameter space into two parts: In one of them instantons exist, while 
in the other instantons do not exist. We further conclude that the one-dimensional line such as defined by eq.(\ref{eq.6-4}) divides the two-dimensional space 
spanned by $c_1$ and $c_{23}$ into the two parts. See Fig.\ref{fig5}.

It is rather difficult to verify this conclusion numerically compared with the case of 2d O(3) sigma model, because it takes a lot of computer time to do it. 
Therefore we have investigated \cite{Iwasaki-Yoshie-1983,Iwasaki-Sakai-Yoshie} the existence of instantons on a lattice with size of $6^4$ with $c_1$ being 
varied and with $c_2=c_3=0$ by the method described in ref.\cite{Iwasaki-Yoshie-1983}. We have found the following: For model $R_3$ ($c_1=-0.3375$) we have 
instantons with topological number $q=1,2,3,4$ for ten random starts, for model IM11 ($c_1=-0.331$) we have instantons with $q=1,2,3$ and for model IM21 
($c_1=-0.293$) we have instantons with $q=2,3,4$. For $\left|c_1\right| < 0.29$, we have no stable instantons on the lattice. Thus the point $c_1=-0.29$ is 
critical. This is consistent with the above conclusion.

One reason in addition to the reason given in section \ref{section6} why we choose model IM11 rather than model IM21 as an improved action for a $8^4$ lattice 
is that we have no instantons with $q=1$ for model IM21 on a $6^4$ lattice as far as we have investigated. We expect that if the size of the lattice is large 
enough, e.g., $16^4$, instantons with $q=1$ also will exist on the lattice for model IM21.

\section{Discussion}
\label{section9}
The general discussion given in section 11 of [I] can be also applied for 4d SU(N) gauge models. We state here only briefly the main points. The improved action 
near the renormalized trajectory has been determined by the perturbation theory in our approach. We are able to calculate non-perturbative effects using this 
action by MC simulations. The effect of instantons, for example, is properly taken into account even on a lattice with small size.

A different approach to improve lattice action is proposed by Symanzik \cite{Symanzik}. The perturbation theory cannot be applied for calculation of physical 
quantities such as the string tension where non-perturbative effects are crucial. It appears that this limitation is not properly taken into consideration in 
Symazik's approach. As shown in section \ref{section6}, the "improved" action obtained by Weisz \cite{Weisz} in Symanzik's approach is not an improved action 
in our criteria.

The continuum limit of a lattice theory is unique and universal for a wide class of lattice action because of the following facts: i) The renormalized trajectory 
is unique and ii) the trajectory for any lattice action approaches asymptotically the renormalized trajectory, if the action belongs to the domain in the parameter 
space which is governed by the non-trivial fixed point. However, it crucially depends on the form of the action how rapidly the trajectory approaches the 
renormalized trajectory. We have given a criterion to estimate this rapidity. From the criterion we can guess the coupling constant where we may expect that 
the expectation value of a physical quantity is identical with that on the renormalized trajectory. The point is not that a physical quantity approximately shows 
a scaling form, but that it is identical with that on the renormalized trajectory.

We would like to calculate various physical quantities with an improved action due to our criteria in addition to the string tension of the SU(3) gauge theory 
\cite{Iwasaki-Yoshie}. We hope that the mass ratios of mesons to baryons will become realistic with an improved action of a relatively small lattice. We also 
conjecture that the first order phase transition \cite{Creutz-Bohr-1981} observed in SU(4) and SU(5) gauge theories with the standard action will not be 
observed with an improved action, because no phase transition will occur on the renormalized trajectory.

We finally argue that Osterwalder-Schrader \cite{Osterwalder} (OS) positivity is satisfied on the renormalized trajectory and consequently it is satisfied, 
at least approximately, for the improved action of 4d SU(N) gauge models, from the same reasoning given in [I].

\begin{acknowledgments}
The numerical calculation has been performed with the FACOM M200 computer at University of Tsukuba and HITAC M200H at KEK. I would like to thank Hirotaka Sugawara 
and other members of KEK for kind hospitality.
\end{acknowledgments}


\clearpage
\appendix

\section{The explicit form of the propagator \cite{Weisz}}
\label{appendixA}
It is easy to show that $D_{\mu\nu}(k)$ may be written in the form
\begin{equation}
\label{eq.A-1}
D_{\mu\nu}(k) = \left(\hat{k}^2\right)^{-2} \left[ \hat{k}_\mu\hat{k}_\nu + 
\sum_\sigma\left(\hat{k}_\sigma\delta_{\mu\nu} - \hat{k}_\nu\delta_{\mu\sigma}\right)A_{\nu\sigma}\hat{k}_\sigma \right],
\end{equation}
with $A_{\mu\nu}$ satisfying (i) $A_{\mu\mu}=0$ for all $\mu$ and (ii) $A_{\mu\nu} = A_{\nu\mu}$.\\
The element $A_{12}$ (and the other elements by appropriate replacement of indices) is given by
\begin{eqnarray}
\label{eq.A-2}
A_{12} = &\frac{1}{\Delta}& 
\Big[\left(\hat{k}^2-\hat{k}^2_2\right)\left(q_{13}q_{14}\hat{k}^2_1 + q_{13}q_{34}\hat{k}^2_3 + q_{14}q_{34}\hat{k}^2_4\right)
\nonumber\\
&+&\left(\hat{k}^2-\hat{k}^2_1\right)\left(q_{23}q_{24}\hat{k}^2_2 + q_{23}q_{34}\hat{k}^2_3 + q_{24}q_{34}\hat{k}^2_4\right)
\nonumber\\
&+& q_{13}q_{24}\left(\hat{k}^2_1+\hat{k}^2_3\right)\left(\hat{k}^2_2+\hat{k}^2_4\right)
+ q_{14}q_{23}\left(\hat{k}^2_1+\hat{k}^2_4\right)\left(\hat{k}^2_2+\hat{k}^2_3\right)
\nonumber\\
&-& q_{12}q_{34}\left(\hat{k}^2_3+\hat{k}^2_4\right)^2
- \left(q_{13}q_{23} + q_{14}q_{24}\right)\hat{k}^2_3\hat{k}^2_4
\nonumber\\
&-&q_{12}\left( q_{13}\hat{k}^2_1\hat{k}^2_4 + q_{14}\hat{k}^2_1\hat{k}^2_3 + q_{23}\hat{k}^2_2\hat{k}^2_4 + q_{24}\hat{k}^2_2\hat{k}^2_3 \right)\Big].
\end{eqnarray}
where
\begin{eqnarray}
\label{eq.A-3}
\Delta &=& \left(\hat{k}^2\right)^{-2} \det D^{-1}
\nonumber\\
&=& \sum_\mu k^4_\mu \prod_{\mu \neq\nu}q_{\mu\nu} + \sum_{\substack{\mu>\nu\\ \rho>\tau\\ (\rho,\tau)\cap(\mu,\nu)=\emptyset}}
{k^2_\mu k^2_\nu q_{\mu\nu}\left(q_{\mu\rho}q_{\nu\tau} + q_{\mu\tau}q_{\nu\rho} \right)}.
\end{eqnarray}

\section{Derivation of eq.(\ref{eq.3-9})}
\label{appendixB}
Let us introduce the following Fourier transformation
\begin{equation}
\label{eq.B-1}
A^{(I)}_\mu(n) = \int_k{e^{i 2^I \left(kn + k_\nu/2\right)}A^{(I)}_\mu(k)},
\end{equation}
because the system for variables $A^{(I)}_\mu(n)$ is that with spacing $2^Ia$. Combining eqs.(\ref{eq.2-6}), (\ref{eq.3-3}) and (\ref{eq.B-1}) we obtain
\begin{equation}
\label{eq.B-2}
A^{(I)}_\mu(k) = e^{-i\left(2^{I-1}-1/2\right)k_\mu}\tilde{H}^{(I)}(k)A^{(0)}_\mu(k),
\end{equation}
where
\begin{equation}
\label{eq.B-3}
\tilde{H}^{(I)}(k) = \prod_{M=0}^{I-1}\frac{1}{8}\prod_{\mu=1}^4\left(e^{i 2^Mk_\mu} + 1\right).
\end{equation}
Let us define the free propagator $D^{(I)}$ for the field $A^{(I)}$ by
\begin{equation}
\label{eq.B-4}
\left<A^{a(I)}_\mu(k)A^{b(I)}_\nu(k')\right> = \delta_{ab}(2\pi)^4\delta^{(4)}(k+k')D^{(I)}_{\mu\nu}(k).
\end{equation}
Then from eq.(\ref{eq.B-2}) we have
\begin{equation}
\label{eq.B-5}
D^{(I)}_{\mu\nu}(k) = e^{-i\left(2^{I-1}-1/2\right)k_\mu}e^{-i\left(2^{I-1}-1/2\right)k_\nu}H^{(I)}(k)D^{(0)}_{\mu\nu}(k),
\end{equation}
where $H^{(I)}(k)$ is defined by eq.(\ref{eq.3-11}) and $D^{(0)}_{\mu\nu}(k) = D_{\mu\nu}(k)$.\\

The expectation value of various Wilson loops may be written as
\begin{equation}
\label{eq.B-6}
W(C) = \sum_{\mu,\nu}{c_{\mu\nu}(k)D_{\mu\nu}(k)}
\end{equation}
for the original system. Then the corresponding expectation value for the $I-th$ system can be written as
\begin{equation}
\label{eq.B-7}
W^{(I)}(C) = \sum_{\mu,\nu}{c_{\mu\nu}(2^Ik)D^{(I)}_{\mu\nu}(k)},
\end{equation}
because the $I-th$ system is that which scaling $2^Ia$. Note that $D_{\mu\nu}(k)$ is odd in $k_\mu$ and $k_\nu$ when $\mu\neq\nu$. Thus from eqs.(\ref{eq.B-5}) 
and(\ref{eq.B-7}) we can derive eq.(\ref{eq.3-9}).

\end{document}